\documentclass[%
 reprint,
 amsmath,amssymb,
 aps,
pra,
]{revtex4-2}

\usepackage{graphicx}
\usepackage{dcolumn}
\usepackage{bm}
\usepackage{color}
\usepackage{amsthm, amssymb, amsfonts}
\usepackage{ulem}
\usepackage{setspace}
\usepackage{cancel}

\usepackage{lipsum, babel}
\usepackage{braket}
\usepackage{mathrsfs}
\usepackage{comment}

\begin{document}

\preprint{APS/123-QED}

\title{Quantum state tomography for Kerr parametric oscillators}

\author{Y. Suzuki$^{1,2}$}
\author{S. Kawabata$^{2,3}$}
\author{T. Yamamoto$^{3,4}$}
\author{S. Masuda$^{2,3}$}
\email{shumpei.masuda@aist.go.jp}
\affiliation{%
$^1$Department of Physics, Faculty of Science Division I, Tokyo University of Science, 1-3 Kagurazaka, Shinjuku-ku, Tokyo 162-8601, Japan
}%
\affiliation{%
$^2$Research Center for Emerging Computing Technologies (RCECT), National Institute of Advanced Industrial Science and Technology (AIST), 1-1-1, Umezono, Tsukuba, Ibaraki 305-8568, Japan
}%
\affiliation{%
$^3$NEC-AIST Quantum Technology Cooperative Research Laboratory, National Institute of Advanced Industrial Science and Technology (AIST), Tsukuba, Ibaraki 305-8568, Japan
}%
\affiliation{%
$^4$System Platform Research Laboratories, NEC Corporation, Kawasaki, Kanagawa 211-0011, Japan
}%





\date{\today}

\begin{abstract}
Kerr parametric oscillators (KPOs) implemented in the circuit QED architecture can operate as qubits.
Their applications to quantum annealing and universal quantum computation have been studied intensely.
For these applications, the readout of the state of KPOs is of practical importance. 
We develop a scheme of state tomography for KPOs with reflection measurement.
Although it is known that the reflection coefficient depends on the state of the KPO,
it is unclear whether tomography of a qubit encoded into a KPO can be performed in a practical way mitigating decoherence during the measurement, and how accurate it is. 
We show that the reflection coefficient has a one-to-one correspondence with a diagonal element of the density matrix of the qubit when a probe frequency is properly chosen and an additional single-photon-drive is introduced.
Thus, our scheme offers a novel way to readout the qubit along an axis of the Bloch sphere, and therefore the reflection measurement and single-qubit gates can constitute state tomography.
\end{abstract}

\maketitle

\section{Introduction}
In the early days of digital computers, classical parametric phase-locked oscillators~\cite{Onyshkevych1959,Goto1959} were utilized as classical bits.
Recently, their quantum counterpart called Kerr parametric oscillators (KPOs) or Kerr-cat qubits~\cite{Milburn1991,Wielinga1993,Goto2016} are attracting much attention in terms of their applications to quantum information processing~\cite{Goto2019} and studies of quantum many-body systems~\cite{Dykman2018,Rota2019}.
A KPO can be implemented by a superconducting resonator with the Kerr nonlinearity, driven by an oscillating pump field~\cite{Meaney2014,Wang2019,Goto2019,Grimm2020}.
In a KPO, two coherent states with opposite phases can exist stably, and be used as qubit states.

An advantage of KPOs as qubits is rooted in a characteristic of their errors. 
The bit-flip error in a KPO is greatly suppressed because of the stability of the coherent states against photon loss.
Thus, the phase-flip error is predominant over the bit-flip error in a KPO.
This biased nature of errors enables us to perform quantum error corrections with less overhead compared to other qubits with unbiased errors~\cite{Tuckett2019,Ataides2021}.

Previous studies on applications of KPOs include theoretical studies of quantum annealing \cite{Goto2016,Nigg2017,Puri2017,Zhao2018,Onodera2020,Goto2020a,Kewming2020,Kanao2021,Yamaji2022}
and universal quantum computation \cite{Cochrane1999,Goto2016b,Puri2017b}, 
experimental demonstration of single-qubit operations~\cite{Grimm2020},
theoretical studies of qubit gate operations~\cite{Puri2020,Kanao2022,Masuda2022,Chono2022,Aoki2023}, 
high error-correction performance by concatenating the XZZX surface code~\cite{Ataides2021} with KPOs~\cite{Darmawan2021}.
There are other subjects such as fast and accurate controls~\cite{Xu2021,Kang2021}, spectroscopy~\cite{Yamaji2021,Masuda2021b}, controls and dynamics not confined in qubit space~\cite{Zhang2017,Wang2019}, Boltzmann sampling~\cite{Goto2018}, effect of strong pump field~\cite{Masuda2020}, effect of decay and dephasing \cite{Puri2017b}, quantum phase transitions~\cite{Dykman2018,Rota2019,Kewming2022} and quantum chaos~\cite{Milburn1991,Hovsepyan2016,Goto2021b}.

In almost all of these applications, the readout of KPOs is essential.
Quantum state tomography of a KPO using the transient power spectrum density (PSD) was demonstrated experimentally in Ref.~\cite{Wang2019}.
In Ref.~\cite{Grimm2020}, authors extracted the state of a KPO by adiabatically transforming it to a Fock qubit and performing state tomography of the Fock qubit. They also proposed a quantum non-demolition measurement along the $z$-axis of the Bloch sphere using an additional readout resonator, where the $z$-axis is defined so that the stable coherent states are located on the axis.

In this paper, we develop a scheme of quantum state tomography with reflection measurement which is widely and routinely used for circuit QED systems.
Our scheme does not require reading a small number of photons in contrast to transient PSD nor a readout resonator.
Also, it does not use a transformation of a KPO to a Fock qubit caused by control of a pump field.
Therefore, it is expected that our scheme can simplify design of KPO systems and avoid decoherence during the adiabatic transformation of a KPO to a Fock qubit.

In Ref.~[\citenum{Masuda2021b}], the reflection coefficient was obtained as a function of density-matrix elements of a KPO, and it was examined especially for a KPO in a stationary state. 
Since the reflection coefficient depends on the state of a  KPO, it is expected that we can estimate the density matrix of a qubit encoded into the KPO with reflection measurements in principle. However, it is not obvious if tomography can be performed  in a practical manner and how accurate it is. 
For example, it is nontrivial which probe frequency should be chosen and what controls are needed.
Because a KPO has effective phase decay rate proportional to the photon number stored in the KPO~\cite{Puri2017b}, the decoherence during the reflection measurement can degrade the efficiency of the tomography.
This paper addresses all of these points.

This paper is organized as follows. In Sec.~\ref{Method of tomography}, our method of state tomography is outlined. In Sec.~\ref{Extraction}, we explain how to extract the diagonal elements of the density matrix of a KPO with reflection measurement. In Sec.~\ref{Sensitivity of reflection coefficient}, we examine the sensitivity of the reflection coefficient to density-matrix elements.
Section~\ref{Summary and discussion} is devoted to summary and discussion.
We present a way to obtain off-diagonal elements and examine the accuracy of the tomography in Appendix~\ref{Fidelity of tomography}.

\section{Method of tomography}
\label{Method of tomography}
\begin{figure}[h]
  \centering
  \includegraphics[width=6.5cm]{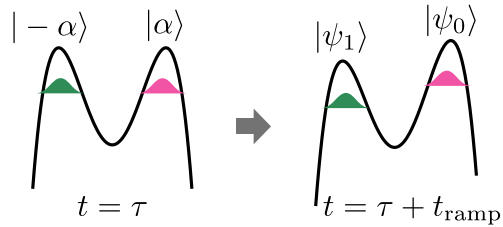}
  \caption{
  Schematic of the effective potential $\bra{\alpha'} H_{\rm KPO} \ket{\alpha'}$, where $\alpha'$ is real.
  The potential is tilted due to the drive amplitude $\Omega$ as explained in Sec.~\ref{Extraction}.
  The pink and green wave packets represent the two highest energy eigenstates. 
  During  the ramp of the drive field, $\ket{\pm \alpha}$ is adiabatically loaded to $\ket{\psi_{0,1}}$, respectively.
}
 \label{poten_sch_10_8_22_2}
\end{figure}

The Hamiltonian of an isolated KPO can be written in a rotating frame at the frequency of $\omega_p/2$ as~\cite{Goto2019}
\begin{eqnarray}
\frac{H_{\rm{KPO}}}{\hbar}=-\frac{K}{2}\hat{a}^{\dagger 2} \hat{a}^2 +\frac{p}{2}\left(\hat{a}^{\dagger 2}+\hat{a}^{2}\right),
\label{Hamiiltonian_not-Omega}
\end{eqnarray}
where $K(>0)$, $p(>0)$ and $\omega_p$ are the nonlinearity parameter, pump amplitude and angular frequency of the pump field, respectively.
The highest and second highest eigenstates of the Hamiltonian (\ref{Hamiiltonian_not-Omega}) are represented as
\begin{eqnarray}
|\varphi_0\rangle &=& N_+ (|\alpha\rangle + |-\alpha\rangle),\nonumber\\
|\varphi_1\rangle &=& N_- (|\alpha\rangle - |-\alpha\rangle),
\end{eqnarray}
where $\ket{\pm\alpha}$ is a coherent state; $\alpha=\sqrt{p/K}$; $N_{\pm}= (2\pm2e^{-2\alpha^2})^{-1/2}$.
It can be verified that these states become Fock states $\ket{0}$ and $\ket{1}$ as $p$ goes to zero, respectively. 
$|\varphi_0\rangle$ and $|\varphi_1\rangle$ can be used as qubit states~\cite{Cochrane1999}.
In this paper we consider the large-$p$ regime where the two coherent states $\ket{\pm \alpha}$ are orthogonal and therefore can be used as qubit states.
We use these coherent states as qubit states.
In this parameter regime, these coherent states are long-lived even with photon loss, and the highest and second highest eigenstates are approximately degenerate.
The coherent states are located around the maxima of the effective potential with an inverted double-well structure (Fig.~\ref{poten_sch_10_8_22_2}).
With these coherent states, the density operator of the KPO at time $\tau$ is represented as
\begin{equation}
\rho(\tau)= \sum_{i,j=0}^1
\rho_{ij}(\tau)\ket{\tilde{i}}\bra{\tilde{j}},
\label{state-alpha}
\end{equation}
where $\ket{\tilde{0}(\tilde{1})} \equiv \ket{\alpha(-\alpha)}$.
In Eq.~(\ref{state-alpha}), we assume that leakage out of the qubit subspace is negligible. 
The condition for this assumption to be valid is reported in Ref.~\cite{Puri2020}.

The purpose of the quantum state tomography is to obtain $\rho_{ij}(\tau)$.
The scheme of the tomography consists of a single-qubit gate and subsequent measurement which determines 
the diagonal elements $\rho_{ii}(\tau)$.
For example, the off-diagonal elements $\rho_{01}(\tau)$ and $\rho_{10}(\tau)$ can be measured with $R_x(\pi/2)$ and $R_y(\pi/2)$ gates followed by the measurement
for diagonal elements.
This is because $R_x(\pi/2)$ and $R_y(\pi/2)$ gates change the coefficient of the term of $\ket{\tilde{0}}\bra{\tilde{0}}$ from $\rho_{00}(\tau)$ to $1/2-\mathrm{Im}[\rho_{01}(\tau)]$ and $1/2-\mathrm{Re}[\rho_{01}(\tau)]$, respectively, as explained in Appendix~\ref{Fidelity of tomography}.

\section{Extraction of $\rho_{ii}$}
\label{Extraction}
Our scheme to extract $\rho_{ii}(\tau)$ can be divided into two processes: 
 ramping of a single-photon-drive field and subsequent reflection measurement.
Hereafter, we refer to the single-photon-drive field as drive field.
The role of the drive field is to make the reflection coefficient dependent on $\rho_{ii}(\tau)$ (see Appendix~\ref{Reflection coefficient without a drive field}). 
At time $t=\tau$, we start to ramp a drive field resonant to the KPO, while $p$ and $K$ are fixed.
All the parameters are kept constant for $t>\tau+t_{\rm ramp}$, where $t_{\rm ramp}$ is the ramping time.
The reflection measurement is started at $t=\tau+t_{\rm ramp}+t_{\rm delay}$, where $t_{\rm delay}$ is the delay time.
We assume that the duration of these processes is sufficiently smaller than the typical time that the bit-flip occurs so that the bit-flip is negligible.
The role of the delay time is explained later.

\subsection{Ramping of drive field}
As the drive field gradually increases, the energy eigenstates $\ket{\tilde{0}}$ and $\ket{\tilde{1}}$ change to $\ket{\psi_0}$ and $\ket{\psi_1}$, respectively, as illustrated in Fig.~\ref{poten_sch_10_8_22_2}. 
Here, $\ket{\psi_{0,1}}$ are the highest and second highest eigenstates of $H_{\rm KPO}(\tau+t_{\rm ramp})$, where 
\begin{eqnarray}
\frac{H_{\rm{KPO}}(t)}{\hbar}&=&-\frac{K}{2}\hat{a}^{\dagger 2} \hat{a}^2 +\frac{p}{2}\left(\hat{a}^{\dagger 2}+\hat{a}^{2}\right)
+ \Omega(t)(\hat{a}^{\dagger} + \hat{a}),\nonumber\\
\label{Hamiiltonian_Omega}
\end{eqnarray}
and $\Omega$ is the drive amplitude.
We set $t_{\rm ramp}$ long enough to suppress undesired non-adiabatic transitions in the state of the KPO.

Then, the diagonal elements of the density operator changes from $\rho_{ii}\ket{\tilde{i}}\bra{\tilde{i}}$ to $\rho_{ii}\ket{\psi_i}\bra{\psi_i}$.
The off-diagonal elements in Eq.~(\ref{state-alpha}) are suppressed much faster than bit-flip.
The effective phase decay rate increases linearly in $|\alpha|^2$ and is explicitly written as $2\kappa_{\rm tot} |\alpha|^2$~\cite{Puri2017b}, where $\kappa_{\rm tot}$ is the total photon loss rate defined by $\kappa_{\rm tot}=\kappa_{\rm ex}+\kappa_{\rm int}$ with the external and internal decay rates, $\kappa_{\rm ex}$ and $\kappa_{\rm int}$.
On the other hand, the bit-flip rate is suppressed exponentially in $|\alpha|^2$~\cite{Puri2019,Suzuki2022}.
The off-diagonal elements vanish when the delay time $t_{\rm delay}$ is long enough.
As a result, the density operator $\rho(t)$ for $t\ge\tau+t_{\rm ramp}+t_{\rm delay}$ is approximated by
\begin{equation}
\begin{split}
\rho'
&=
\rho_{00}(\tau)\ket{\psi_0}\bra{\psi_0}+\rho_{11}(\tau)\ket{\psi_1}\bra{\psi_1}.
\end{split}
\label{after_relaxation}
\end{equation}
In this manner, our method takes advantage of the biased nature of errors of the KPO: the typical time that the phase-flip  occurs is much shorter than the typical time of the bit-flip.


In order to demonstrate the time evolution of a KPO discussed above, we numerically solve the master equation in the Lindbladian form expressed as 
\begin{equation}
\begin{split}
\frac{d\rho(t)}{dt}
&=
-\frac{i}{\hbar}
\left[H_{\rm KPO}(t),\rho(t)\right]
+
\frac{\kappa_{\rm tot}}{2}
\mathcal{D}[\hat{a}]\rho(t),
\end{split}
\label{master eq}
\end{equation}
where the Lindbladian superoperator term is defined by $\mathcal{D}[\hat{O}]\rho=2\hat{O}\rho\hat{O}^\dagger-\hat{O}^\dagger\hat{O}\rho-\rho \hat{O}^\dagger\hat{O}$.
In the numerical simulation, we set $\tau=0$ and  $\rho(0)=(\sqrt{0.2}\ket{\tilde{0}} + \sqrt{0.8}\ket{\tilde{1}})(\sqrt{0.2}\bra{\tilde{0}} + \sqrt{0.8}\bra{\tilde{1}})$.
The time dependence of the drive amplitude is given by
\begin{eqnarray}
\Omega(t) = \frac{\Omega_0}{2}\Big{[} 1 - \cos\Big{(} \frac{\pi t}{t_{\rm ramp}}\Big{)} \Big{]}
\label{Omega_11_15_22}
\end{eqnarray}
for $0\le t\le t_{\rm ramp}$ so that $\Omega$ is gradually increased.
The time dependence of $\Omega$ in Eq.~(\ref{Omega_11_15_22}) was chosen so that $\Omega$ and $d\Omega/dt$ are continuous to avoid unwanted nonadiabatic transitions. 

\begin{figure}[h]
  \centering
  \includegraphics[width=8.6cm]{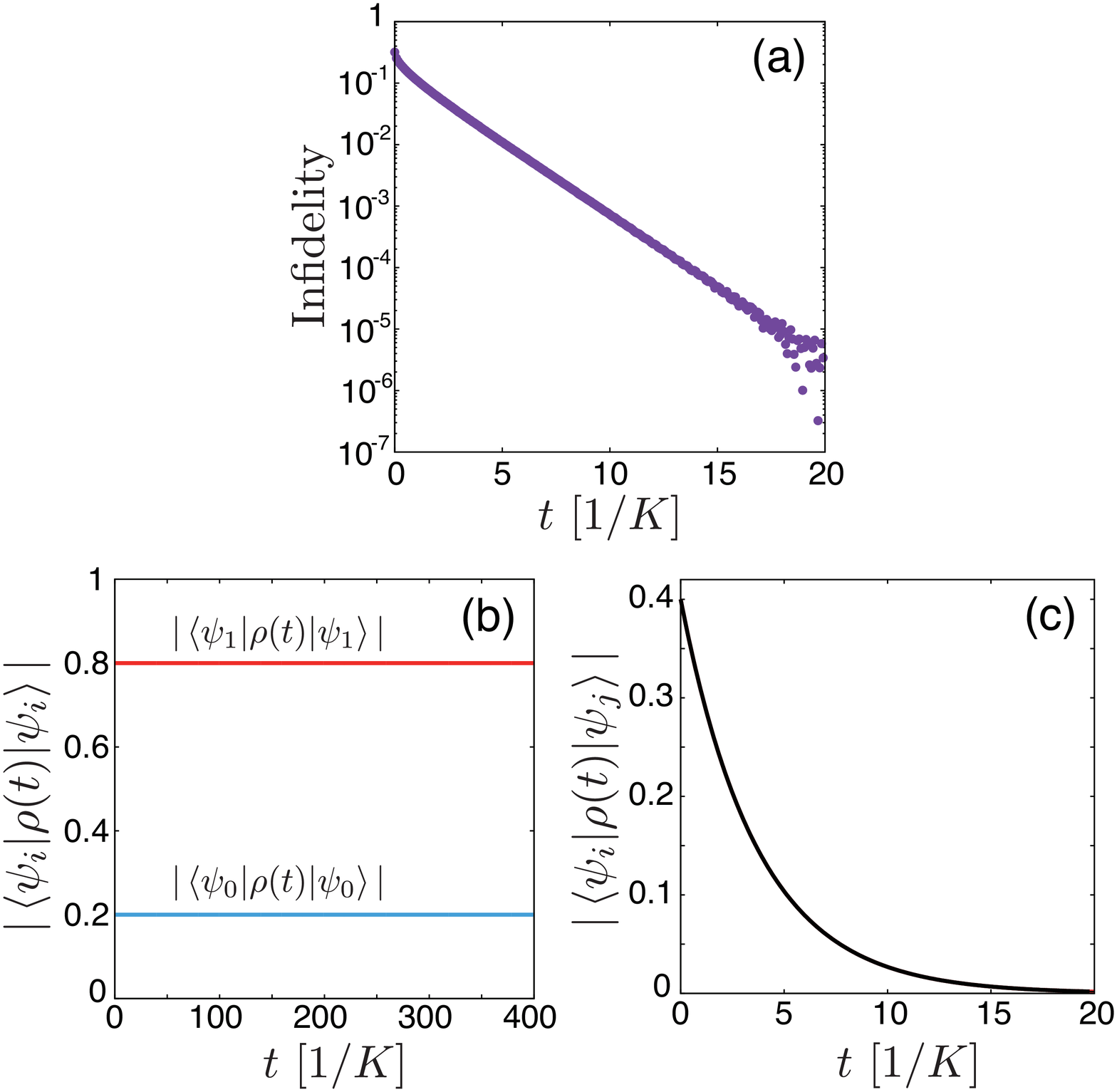}
  \caption{
(a) Infidelity in logarithmic scale, $1-\mathcal{F}[\rho(t),\rho']$, as a function of $t$.
The amplitude of diagonal (b) and off-diagonal elements (c) of the density matrix as a function of $t$. 
The blue and red curves in panel (b) and the black curve in panel (c) represent $|\bra{\psi_0}\rho(t)\ket{\psi_0}|$, $|\bra{\psi_1}\rho(t)\ket{\psi_1}|$ and $|\bra{\psi_{0(1)}}\rho(t)\ket{\psi_{1(0)}}|$, respectively.
The used parameters are $p/K=9.0$, $t_{\rm ramp}=20/K$, $\Omega_0/K=0.1$, $\kappa_{\rm ex}/K=0.01$ and $\kappa_{\rm in}/\kappa_{\rm ex}=0.5$.
}
 \label{fid_exp}
\end{figure}

We consider the fidelity between $\rho(t)$ and $\rho'$ in Eq.~(\ref{after_relaxation}) defined by $\mathcal{F}[\rho(t),\rho']$, where $\mathcal{F}[\rho,\rho']=({\rm Tr}[\sqrt{\sqrt{\rho}\rho'\sqrt{\rho}}])^2$.
Figure~\ref{fid_exp}(a) shows the infidelity defined by $1-\mathcal{F}[\rho(t),\rho']$. 
The infidelity tends to decrease with the increase of $t$, and is lower than $10^{-3}$ for $t\ge 10/K$.
Figures~\ref{fid_exp}(b) and \ref{fid_exp}(c) represent the diagonal and off-diagonal elements of the density matrix as a function of $t$, respectively.
The diagonal elements are almost unchanged for $0<t<400/K$, while the off-diagonal elements vanish rapidly.
These results indicates that $\rho'$ approximates well $\rho(t)$ for a considerably long period of time.

\subsection{Reflection measurement}
\begin{figure}[]
  \centering
  \includegraphics[width=6cm]{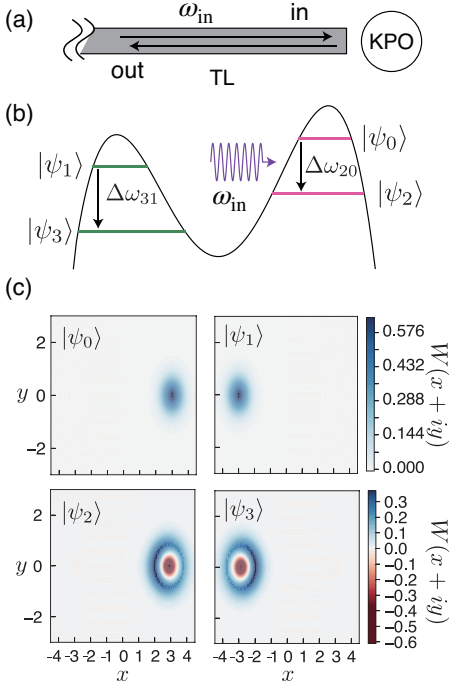}
  \caption{
  (a) Schematic of reflection measurement.
  A microwave with frequency of $\omega_{\rm in}$ is injected from a transmission line (TL) attached to the KPO.
  The reflected microwave is measured.
  (b)  Schematic of the effective potential tilted due to the drive amplitude $\Omega$.
  The straight arrows indicate transitions $\ket{\psi_0}\rightarrow \ket{\psi_2}$ and $\ket{\psi_1}\rightarrow \ket{\psi_3}$ caused by the input field.
  Here, $\Delta\omega_{nm}$ is the energy difference between $\ket{\psi_n}$ and $\ket{\psi_m}$ defined by $\Delta\omega_{nm}=\omega_n-\omega_m$, where $\omega_m$ is the eigenenergy of $\ket{\psi_m}$.
  The oscillating curve with an arrow indicates the input field with frequency $\omega_{\rm in}$.
  (c) Wigner function, $W(x+iy)$, of the four highest energy levels for $p/K=9$ and $\Omega/K=0.1$.
  }
 \label{equipment}
\end{figure}

We consider reflection measurement of the KPO subjected to a fixed pump and drive fields at $t=\tau+t_{\rm ramp}+t_{\rm delay}$ [Fig.~\ref{equipment}(a)].
A microwave with frequency of $\omega_{\rm in}$ is injected from a TL attached to the KPO.

A recent study~\cite{Masuda2021b} finds that the reflection coefficient is represented as
\begin{equation}
\begin{split}
\Gamma
=
1+\sum_{mn}\xi_{mn}
\end{split}
\label{reflection_coefficient}
\end{equation}
with
\begin{equation}
\begin{split}
\xi_{mn}
=
\frac
{\kappa_{\rm ex} X_{mn}\sum_{k}\left(X^\ast_{kn}\rho^{({\rm F})}_{km}[0]-\rho^{({\rm F})}_{nk}[0] X^\ast_{mk}\right)}
{
i\Delta_{nm}
+\kappa_{\rm tot}X_{nn}X^\ast_{mm}
-\frac{\kappa_{\rm tot}}{2}\left(Y_{nn}+Y_{mm}\right)
},
\\
\end{split}
\label{Gamma_mn}
\end{equation}
where $\Delta_{nm}=\omega_{\rm in}-\omega_p/2-\omega_{n}+\omega_{m}$.
In Eq.~(\ref{Gamma_mn}), $X_{mn}=\braket{\psi_m|\hat{a}|\psi_n}$ and $Y_{mn}=\braket{\psi_m|\hat{a}^\dagger \hat{a}|\psi_n}$. 
$\rho^{({\rm F})}_{mn}[0]$ is the Fourier component of $\braket{\psi_m|\rho|\psi_n}$ at frequency of $0$ in the rotating frame. 
Thus, the information of the density operator is embedded in the reflection coefficient.
In Eq.~(\ref{reflection_coefficient}), $\xi_{mn}$ can be interpreted as the contribution from the transition $\ket{\psi_m}\rightarrow\ket{\psi_n}$ to the reflection coefficient. 
When the input field is resonant with this transition, that is, $\omega_{\rm in}-\omega_p/2=\omega_n-\omega_m$,  the amplitude of $\xi_{mn}$ becomes large because its denominator becomes small.
Thus, the resonant  transitions dominate the reflection coefficient over the other non-resonant transitions.

We assume that the reflection measurement is performed after the off-diagonal elements of the density matrix vanish, and that bit-flip does not occur during the measurement.
If the input field is sufficiently weak, the change in $\rho^{({\rm F})}_{mn}[0]$ from that given by Eq.~(\ref{after_relaxation}) is negligible.
Therefore, we approximately have
\begin{eqnarray}
\rho^{({\rm F})}_{mn}[0]=\rho_{00}(\tau)\delta_{m0}\delta_{n0} +(1-\rho_{00}(\tau))\delta_{m1}\delta_{n1}, 
\label{Pmn_9_29_22}
\end{eqnarray}
where we used $\rho_{11}(\tau)=1-\rho_{00}(\tau)$.
Straightforward substitution of Eqs.~(\ref{Gamma_mn}) and (\ref{Pmn_9_29_22}) into Eq.~(\ref{reflection_coefficient}) shows a linear relationship between $\Gamma$ and $\rho_{00}(\tau)$.
This linear relationship suggests that measurement of $\Gamma$ allows to extract $\rho_{00}(\tau)$.
Hereafter, we write the reflection coefficient as $\Gamma[\rho_{00}(\tau)]$ to  clearly  express that it  depends on  $\rho_{00}(\tau)$.

In this paper, we mainly consider the case that the off-diagonal elements $\rho^{({\rm F})}_{mn(\ne m)}[0]$ are zero during the reflection measurement for simplicity.
However, the effect of the off-diagonal elements is actually negligible when the pump amplitude is sufficiently large as shown in Appendix~\ref{Effect of off-diagonal elements of the density matrix}.
Therefore, $\tau_{\rm delay}$ can be set to zero in such a parameter regime.

\section{Sensitivity of reflection coefficient}
\label{Sensitivity of reflection coefficient}
Accurate extraction of $\rho_{00}(\tau)$ requires sufficient sensitivity of the reflection coefficient to $\rho_{00}(\tau)$.
In this section, we consider $|\Gamma(1)-\Gamma(0)|$ as a measure of the sensitivity and 
show that high sensitivity is obtained with experimentally feasible parameters.
Hereafter, we call $|\Gamma(1)-\Gamma(0)|$ sensitivity.

\subsection{Analytic formula in large-pump limit}
We derive an asymptotic formula of the sensitivity in the large-pump limit.
As an example, we consider the case that the input field is resonant with the transition $\ket{\psi_0}\rightarrow\ket{\psi_2}$ and off-resonant with the other transitions [Fig.~\ref{equipment}(b)].  
Then, the reflection coefficient is approximately given by $\Gamma=1+\xi_{02}$.
When $p$  is sufficiently large, there is a regime of $\Omega$, where the highest levels are well approximated by $D(\pm\alpha)\ket{m}$~\cite{Wang2019} with the displacement operator  $D(\alpha)=\exp\left(\alpha \hat{a}^\dagger-\alpha^*\hat{a}\right)$ (see Appendix~\ref{Energy eigenstates}). 
The Wigner function of the four highest energy levels are exhibited in Fig.~\ref{equipment}(c).
The Wigner function is defined by $W(z)=2\mathrm{Tr}[D(z)\rho D(-z)P]/\pi$ with $z=x+iy$ 
with the parity operator $P=\exp(i \pi \hat{a}^\dagger \hat{a})$.
The use of $\ket{\psi_{0(2)}}=D(\alpha)\ket{0(1)}$ and Eq.~(\ref{Gamma_mn}) leads to
\begin{equation}
|\Gamma(1)-\Gamma(0)|=\frac{2\kappa_{\rm ex}}{\kappa_{\rm tot}}.
\label{Gamma_10_11_22}
\end{equation}
The same result can be obtained for the input field resonant with the transition $\ket{\psi_1}\rightarrow\ket{\psi_3}$.
The above discussion suggests that the sensitivity approaches two, which is the maximum value of the sensitivity, when the pump amplitude becomes strong and $\kappa_{\rm int} \ll \kappa_{\rm ex}$. 
This large sensitivity of the reflection coefficient to $\rho_{00}(\tau)$ is useful for extraction of $\rho_{00}(\tau)$.

\subsection{Numerical results}
\begin{figure}[]
  \centering
  \includegraphics[width=8.6cm]{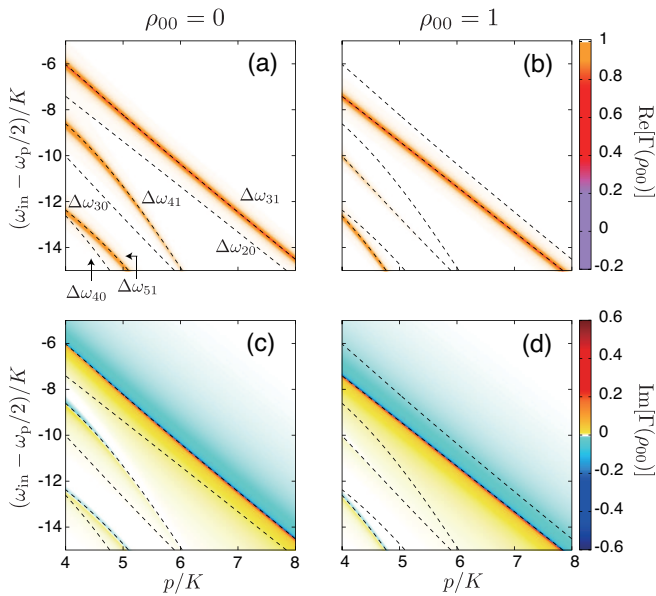}
  \caption{
Real part (a,b) and imaginary part (c,d) of the reflection coefficient $\Gamma$ as function of  $\omega_{\rm in}$ and $p$ for $\rho_{00}=0$ and $1$.
The used parameters are $\Omega/K=0.5$, $\kappa_{\rm ex}/K=0.01$, $\kappa_{\rm in}/\kappa_{\rm ex}=0.5$.
}
 \label{color-map}
\end{figure}

The reflection coefficients for $\rho_{00}(\tau)=0$ and $1$ are compared in Fig.~\ref{color-map}.
The reflection coefficient clearly changes with the value of $\rho_{00}$.
Especially, the difference between $\Gamma(1)$ and $\Gamma(0)$ is large 
for $\omega_{\rm in}-\omega_p/2=\Delta \omega_{20}$ and $\Delta \omega_{31}$ corresponding to the transitions $\ket{\psi_0}\rightarrow\ket{\psi_2}$ and $\ket{\psi_1}\rightarrow\ket{\psi_3}$, where $\Delta \omega_{nm}=\omega_n-\omega_m$.
Therefore, these values of $\omega_{\rm in}$ is suitable for extraction of $\rho_{00}$.

\begin{figure}[t]
  \centering
  \includegraphics[width=8.6cm]{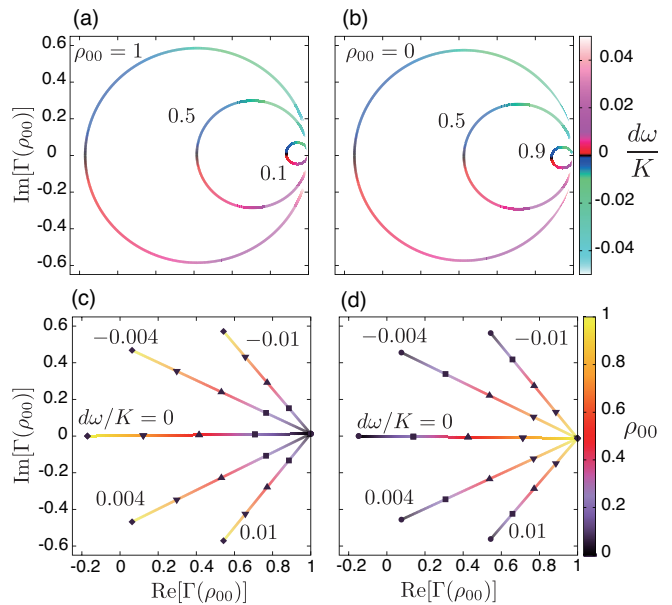}
  \caption{
Reflection coefficient in a complex plane for varying $\omega_{\rm in}$~(a,b) and  fixed $\omega_{\rm in}$~(c,d).
For panels~(a,c) and~(b,d), $d\omega$ is defined by $d\omega=\omega_{\rm in}-\omega_{p}/2-\Delta\omega_{20}$ and $d\omega=\omega_{\rm in}-\omega_{p}/2-\Delta\omega_{31}$, respectively.
The used values of $\rho_{00}$ and $d\omega$ are written on panels (a,b) and (c,d), respectively.  
In panels (c,d), the black circle, square, triangle, inverted triangle, and diamond, are for $\rho_{00}=0.0$, $0.25$, $0.5$, $0.75$, and $1.0$, respectively.
The used parameters are $\Omega/K=0.5$, $p/K=9.0$, $\kappa_{\rm ex}/K=0.01$ and $\kappa_{\rm in}/\kappa_{\rm ex}=0.5$.
}
 \label{complex-plane}
\end{figure}

Figure~\ref{complex-plane} shows the reflection coefficient in a complex plane for varying $\omega_{\rm in}$ (a,b) and  fixed $\omega_{\rm in}$  (c,d).
The numerical results are consistent with the linearity and sensitivity to $\rho_{00}$ proven analytically.
Because there is a one-to-one correspondence between $\rho_{00}$ and $\Gamma$,
$\rho_{00}$ can be determined from $\Gamma$.

\begin{figure}[h]
  \centering
  \includegraphics[width=7cm]{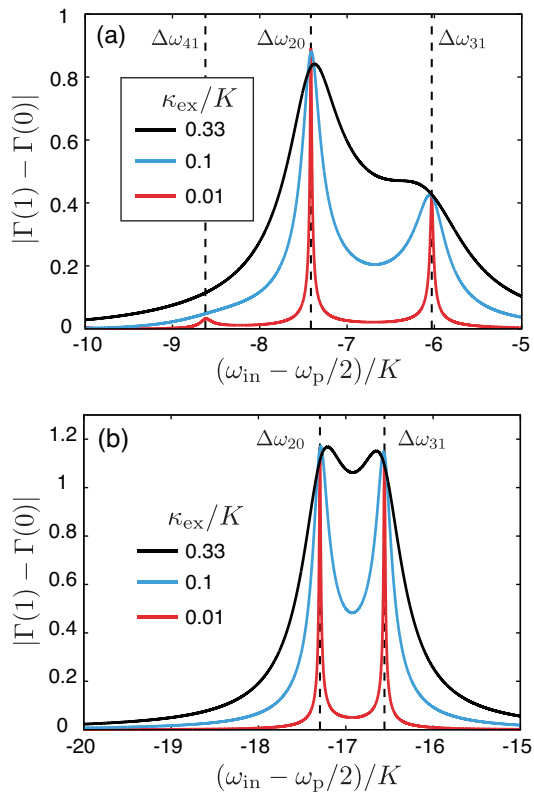}
  \caption{
  Sensitivity as a function of $\omega_{\rm in}$ for $p/K=4.0$~(a) and $p/K=9.0$~(b).
  Black, blue and red curves are for $\kappa_{\rm ex}/K=0.33$, $0.1$ and $0.01$.
  We also vary $\kappa_{\rm int}$ so that $\kappa_{\rm int}/\kappa_{\rm ex}=0.5$.
  Black dashed lines represent the energy differences between relevant levels.
  We set $\Omega/K=0.5$.
}
 \label{beta20_45}
\end{figure}

Figure~\ref{beta20_45} shows the sensitivity as a function of $\omega_{\rm in}$ for different  values of $\kappa_{\rm ex}$ and $p$. 
There are high peaks of the sensitivity at $\omega_{\rm in}$ corresponding to the transitions $\ket{\psi_0}\rightarrow\ket{\psi_2}$ and $\ket{\psi_1}\rightarrow\ket{\psi_3}$.
There is also a small peak corresponding to the transition $\ket{\psi_1}\rightarrow\ket{\psi_4}$ when $\kappa_{\rm ex}$ and $p$ are small.
This peak is buried in a higher adjacent peak when $\kappa_{\rm ex}$ becomes large as each peak becomes broader.

\begin{figure}[h]
  \centering
  \includegraphics[width=7cm]{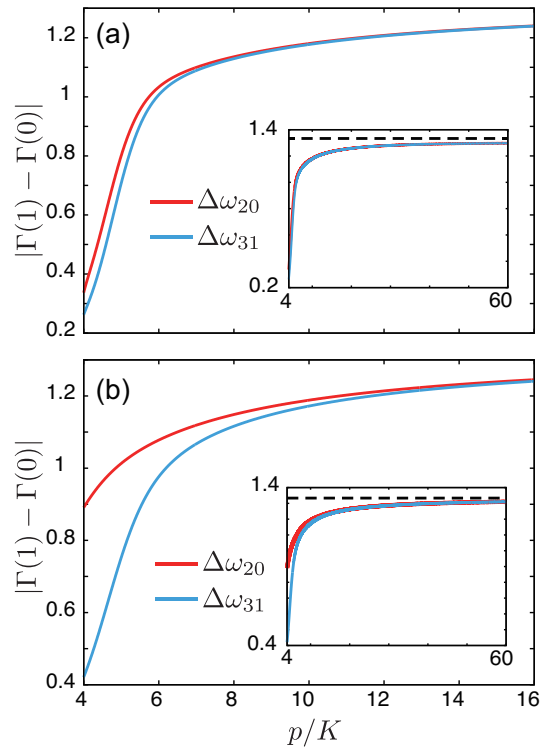}
  \caption{
The sensitivity as a function of $p$ for $\Omega/K=0.1$~(a) and 0.5~(b).
The red and blue curves are for $\omega_{\rm in}-\omega_{p}/2=\Delta\omega_{20}$ and  $\Delta\omega_{31}$, respectively.
The insets show the same things but with a wider range of $p$.
The red and blue curves are overlapping in the inset of panel (a).
The used parameters are $\kappa_{\rm ex}/K=0.01$ and $\kappa_{\rm in}/\kappa_{\rm ex}=0.5$.
The asymptotic value of the sensitivity in Eq.~(\ref{Gamma_10_11_22}) is 1.33 approximately, and is represented by dashed lines in the insets.
}
 \label{01_and_05_10_21_22}
\end{figure}

The sensitivity depends on the pump amplitude.
Figure~\ref{01_and_05_10_21_22} shows the dependence of the sensitivity on the pump amplitude $p$.
The frequency of the input field is set as $\omega_{\rm in}-\omega_{p}/2=\Delta\omega_{20}$ or  $\Delta\omega_{31}$. 
The sensitivity increases with $p$ toward the asymptotic value in Eq.~(\ref{Gamma_10_11_22}).
This is because that relevant energy eigenstates are approximated well by $D(\pm\alpha)\ket{m}$ when $p$ becomes large as numerically confirmed in Appendix~\ref{Energy eigenstates}.
The sensitivity corresponding to the transition $\ket{\psi_0}\rightarrow \ket{\psi_2}$ is higher than that for $\ket{\psi_1}\rightarrow \ket{\psi_3}$. 
We attribute this to that $\ket{\psi_{2}}$ is approximated well by $D(\alpha)\ket{1}$ while $\ket{\psi_{3}}$
deviates from $D(-\alpha)\ket{1}$ especially when $p$ is small as numerically demonstrated in Appendix~\ref{Energy eigenstates}.
Because the energy of $\ket{\psi_3}$ is lower than that of $\ket{\psi_2}$,
$\ket{\psi_3}$ is more loosely trapped than $\ket{\psi_2}$ in one of the potential wells of the inverted double-well potential.

\begin{figure}[h]
  \centering
  \includegraphics[width=7cm]{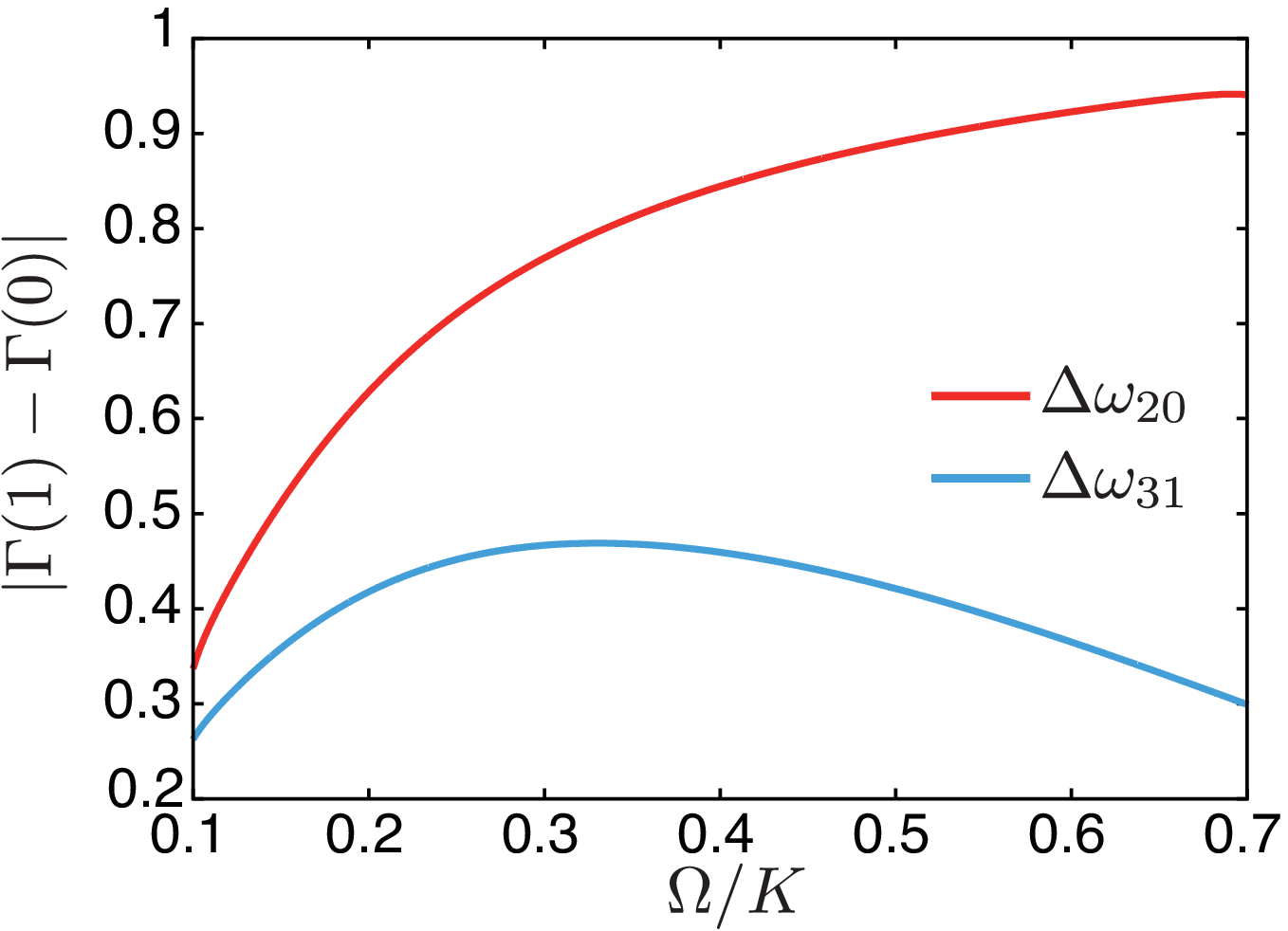}
  \caption{
The sensitivity as a function of $\Omega$ for $p/K=4.0$.
The red and blue curves are for $\omega_{\rm in}-\omega_{p}/2=\Delta\omega_{20}$ and  $\Delta\omega_{31}$, respectively.
The used parameters are $\kappa_{\rm ex}/K=0.01$ and $\kappa_{\rm in}/\kappa_{\rm ex}=0.5$.
}
 \label{2_and_3_10_21_22}
\end{figure}

The sensitivity depends also on the drive amplitude.
Figure~\ref{2_and_3_10_21_22} shows the sensitivity as a function of $\Omega$.
The sensitivity corresponding to the transition $\ket{\psi_1}\rightarrow \ket{\psi_3}$ decreases for large $\Omega$ regime, while the sensitivity corresponding to the transition $\ket{\psi_0}\rightarrow \ket{\psi_2}$ increases monotonically in the range of $\Omega$ used.
The decrease of the sensitivity is due to the deviation of $\ket{\psi_3}$ from $D(-\alpha)\ket{1}$, which is discussed in Appendix~\ref{Energy eigenstates}.
Because $\ket{\psi_3}$ is more loosely trapped than $\ket{\psi_2}$, the state vector of $\ket{\psi_3}$ is sensitive to $\Omega$ compared to $\ket{\psi_2}$.
When $\Omega$ increases greater than $0.7K$, the order of energy levels is changed.
For example, the energy level approximated by $D(-\alpha)\ket{0}$ becomes lower than the level approximated by $D(\alpha)\ket{1}$.
We do not consider such regime of $\Omega$ for simplicity.
When the input field is resonant to different transitions, the formula of the reflection coefficient in Eq.~(\ref{reflection_coefficient}) becomes less valid due to interference between the transitions~\cite{Masuda2021b}. We do not consider such parameter regimes in this paper.

So far, we explained the method to obtain the diagonal elements of the density matrix, $\rho_{00}(\tau)$ and $\rho_{11}(\tau)$. 
In Appendix~\ref{Fidelity of tomography}, we explain the method to obtain the off-diagonal elements, $\rho_{01}(\tau)$ and $\rho_{10}(\tau)$, and examine the accuracy of the tomography.

\section{Summary and discussion}
\label{Summary and discussion}
We have developed a scheme of quantum state tomography for KPOs with reflection measurement.
A drive field is used to realize the one-to-one correspondence between the reflection coefficient and the diagonal elements of the density matrix.
We have examined the sensitivity of the reflection coefficient to the density matrix for various system parameters. In addition, we have examined the accuracy of the tomography by numerically simulating the gate operations.
Our scheme does not use readout of a small number of photons in contrast to transient PSD.
Moreover, it does not require a transformation of a KPO to a Fock qubit nor an additional resonator.
This measurement along $z$-axis can be used also for readouts of KPOs in quantum annealing.

The proposed scheme utilizes the direct coupling of a KPO to a TL. 
However, the external coupling to the TL degrades the fidelity of gate operations because of the effective phase decay rate increasing with $\kappa_{\rm tot}$ (see Appendix~\ref{Fidelity of tomography} for the accuracy of the tomography under the effect of the effective phase decay). Therefore, $\kappa_{\rm ex}$ should be sufficiently small although a part of the reflected field $\xi_{mn}$ in Eq.~(\ref{Gamma_mn}) containing the information of the KPO becomes weak as it is proportional to $\kappa_{\rm ex}$.
The weakness of the signal can be compensated by the data accumulation time of the reflection measurement and the number of ensemble average.

The pure dephasing if there exists can degrade the efficiency of the measurement.
The degradation is rooted not only in the enhanced relaxation of a KPO but also in the large nominal internal decay rate which brings the KPO to an under-coupling regime.
The effect of the pure dephasing to the reflection coefficient is examined in Appendix~\ref{Pure dephasiing}.

Although our scheme requires only three types of gate operations, $I$, $R_x(\pi/2)$, $R_y(\pi/2)$, to extract the density-matrix elements,
other gate operations can be additionally used to further increase the accuracy of the tomography. Then, the numerical optimization techniques such as those used in Ref.~[\citenum{Wang2019}] will be useful to find an estimated density matrix which reproduces well measurement results.

Continuous homodyne and heterodyne measurements along the $z$-axis by reading the leaked field from a KPO were theoretically studied~\cite{Bartolo2017,Suzuki2022} and experimentally demonstrated~\cite{Yamaji2021}.
These measurements are continuous in the sense that the leaked field is kept measured while a pump field is on.
Homodyne and heterodyne measurement could be alternatively used for exstraction of diagonal elements of the density matrix if the timing of the detection is controlled.

\begin{acknowledgments}
It is a pleasure to acknowledge discussions with T. Nikuni, M. Kunimi, T. Yamaji, A. Yamaguchi and T. Ishikawa.
This paper is partly based on results obtained from a project, JPNP16007, commissioned by the New Energy and Industrial Technology Development Organization (NEDO), Japan. 
\end{acknowledgments}

\appendix

\section{Off-diagonal elements and accuracy of tomography}
\label{Fidelity of tomography}
The off-diagonal elements of the density matrix, $\rho_{01}(\tau)$ and $\rho_{10}(\tau)$ can be measured with $R_x(\pi/2)$ and $R_y(\pi/2)$ gates followed by the measurement
for diagonal elements.
The gate operations transform $\rho(\tau)$ in Eq.~(\ref{state-alpha}) as
\begin{eqnarray}
&&R_x{(\pi/2)}
\rho(\tau)
R_x{(\pi/2)}^\dagger \nonumber\\
&&=
\frac{1}{2}
\begin{pmatrix}
  1-2\rm{Im}[\rho_{01}] & i\rho_{00}+\rho_{10}+\rho_{01}-i\rho_{11} \\
   -i\rho_{00}+\rho_{10}+\rho_{01}+i\rho_{11} & 1+2\rm{Im}[\rho_{01}]
\end{pmatrix},
\nonumber\\
\label{Rx_3_20_23}
\\
&&R_y{(\pi/2)}
\rho(\tau)
R_y{(\pi/2)}^\dagger \nonumber\\
&&=
\frac{1}{2}
\begin{pmatrix}
  1-2\rm{Re}[\rho_{01}] & \rho_{00}+\rho_{10}-\rho_{01}-\rho_{11} \\
   \rho_{00}-\rho_{10}+\rho_{01}-\rho_{11} & 1+2\rm{Re}[\rho_{01}]
\end{pmatrix},
\label{rho_gate_3_19_22}
\end{eqnarray}
where $\rho_{ij}$ abbreviates $\rho_{ij}(\tau)$, and $R_x{(\theta)}$ and  $R_y{(\theta)}$ are defined by
\begin{eqnarray}
R_x(\theta) = \begin{pmatrix}
  \cos(\theta/2) & -i \sin(\theta/2) \\
   -i \sin(\theta/2) & \cos(\theta/2)
\end{pmatrix}, \nonumber\\
R_y(\theta) = \begin{pmatrix}
  \cos(\theta/2) & - \sin(\theta/2) \\
   - \sin(\theta/2) & \cos(\theta/2)
\end{pmatrix}.
\end{eqnarray}
As seen from Eqs.~(\ref{Rx_3_20_23}) and (\ref{rho_gate_3_19_22}), the measurement for diagonal elements of the density matrix after the gate operations allows us to extract $\rho_{01}(\tau)$ and $\rho_{10}(\tau)$.
We hereafter put  $\tau$ to be zero for simplicity of notation.

The $R_x$ gate for a KPO can be implemented by temporally controlling the detuning~\cite{Goto2016b}, which is the difference between the resonance frequency of the KPO and half the frequency of the pump field.  
The role of the detuning is to lift the degeneracy between the two highest levels and to imprint the different dynamical phase to each level which gives rise to the $R_x$ gate.
The Hamiltonian of the KPO with detuning $\Delta$ is written as
\begin{eqnarray}
\frac{H_{\rm{KPO}}(t)}{\hbar}&=&\Delta(t)a^\dagger a
-\frac{K}{2}\hat{a}^{\dagger 2} \hat{a}^2 
+\frac{p}{2}\left(\hat{a}^{\dagger 2}+\hat{a}^{2}\right).
\end{eqnarray}
The time dependence of the detuning is chosen as
\begin{eqnarray}
\Delta(t) = \Delta_0 \sin^2\frac{\pi t}{T_x},
\end{eqnarray}
for $0 \le t\le T_x$. 
The value of $\Delta_0$ is chosen so that the gate fidelity is maximized for given $T_x$, $p$, $K$, and $\theta$ in the case without decoherence. 
For example, $\Delta_0=-6.938K$ is used to obtain the gate fidelity of 0.997 for $T_x=2.5/K$,  $p/K=9$ and $\theta=\pi/2$ in the case without decoherence.
Thus, we have $\rho(T_x)\simeq R_x(\pi/2)\rho(0)R_x(\pi/2)^\dagger$.
By using $\rho(T_x) = R_x(\pi/2)\rho(0)R_x(\pi/2)^\dagger$ and Eq.~(\ref{Rx_3_20_23}), we obtain
 $\rho_{00}(T_x) = 1/2 - {\rm Im}[\rho_{01}(0)]$ and  $\rho_{11}(T_x) = 1/2 + {\rm Im}[\rho_{01}(0)]$.
Therefore, we  can obtain ${\rm Im}[\rho_{01}(0)]$ by measuring  
the diagonal elements of $\rho(T_x)$ with the method introduced in Sec.~\ref{Extraction}.

Because $R_y(\theta)=R_x(-\pi/2)R_z(\theta)R_x(\pi/2)$, we can use  two $R_x$ gates and an $R_z$ gate to realize an $R_y(\theta)$ gate.
It is known that the $R_z$ gate of a KPO can be implemented by a pulsed drive field~\cite{Goto2016b}.
The Hamiltonian is given by Eq.~(\ref{Hamiiltonian_Omega}).
The time dependence of $\Omega$ for an $R_z(\theta)$ gate is given by
\begin{eqnarray}
\Omega(t) = \frac{\pi \theta}{8T_z\sqrt{p/K}} \sin\Big{(} \frac{\pi (t-T_x)}{T_z}\Big{)}
\end{eqnarray}
for $T_x \le t \le T_x + T_z$, where $T_x$ and $T_z$ are the duration of an $R_x$ gate and an $R_z$ gates, respectively.
Assuming $\rho(2T_x + T_z) = R_y(\pi/2)\rho(0)R_y(\pi/2)^\dagger$, we obtain
 $\rho_{00}(2T_x + T_z) = 1/2 - {\rm Re}[\rho_{01}(0)]$ and  $\rho_{11}(2T_x + T_z) = 1/2 + {\rm Re}[\rho_{01}(0)]$.
Therefore, we  can obtain ${\rm Re}[\rho_{01}(0)]$ by measuring  
the diagonal elements of the density matrix after these gate operations.

We simulate the evolution of the system during gate operations and the ramp of the drive field and the free evolution for $t_{\rm delay}$, by numerically integrating the master equation taking into account the effect of $\kappa_{\rm ex}$.
The diagonal elements of the density matrix at $t_f\equiv T_g + t_{\rm ramp} + t_{\rm delay}$ are used to reconstruct $\rho(\tau=0)$, where $T_g$ is the duration of gate operations.
Because no gate operation is required, $T_g$ is zero for extraction of the diagonal elements of $\rho(0)$.
We set $t_{\rm delay}=0.4 /\kappa_{\rm ex}$ so that $t_{\rm delay}$ is inversely proportional to $\kappa_{\rm ex}$ because vanishing of the off-diagonal elements takes longer time for smaller total photon loss rate.
We assume that the diagonal elements of the density matrix at $t=t_f$ can be extracted exactly with the reflection measurement. The effect of the imperfection of the extraction of the diagonal elements is discussed later.

\begin{figure}[h]
  \centering
  \includegraphics[width=7cm]{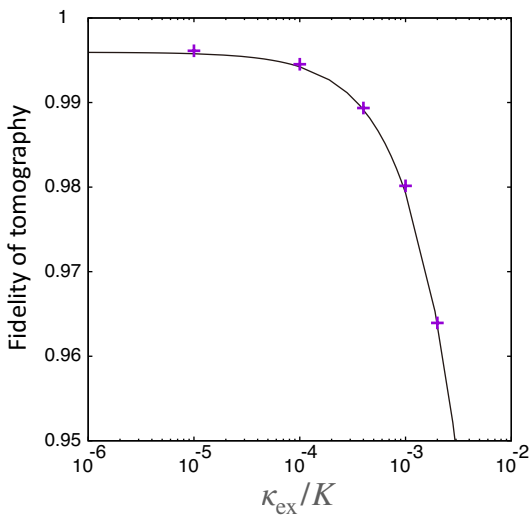}
  \caption{
The fidelity of the tomography as a function of $\kappa_{\rm ex}$ (crosses).
The solid curve is obtained with the approximation
that the diagonal elements of the density matrix at $t=t_f$ are the same as those at $t=T_g$.
The fidelity is averaged over the six different reference states.
The used parameters are $p/K=9.0$, $T_x=2.5/K$, $T_z=1/K$ and $\kappa_{\rm int}=\kappa_{\rm ex}/2$.
}
 \label{Fidelity_Rxy_3_18_23}
\end{figure}

As an example, we consider six different reference states,
$\rho_j = |j\rangle \langle j|$, at $t=0$
where $j$ denotes $\{x\pm, y\pm, z\pm \}$; $|x\pm\rangle = C_{x\pm} (|\tilde{0}\rangle \pm   |\tilde{1}\rangle)$, $|y\pm \rangle = C_{y\pm} (|\tilde{0}\rangle \pm i   |\tilde{1}\rangle)$ and  $|z+(-) \rangle =  |\tilde{0}(\tilde{1})\rangle$;  $C_{j}$ is a normalization factor.
The fidelity of the tomography is defined by $\mathcal{F}[\rho_j,\rho_j']$, where $\rho_j'$ is the reconstructed density operator. 
Figure~\ref{Fidelity_Rxy_3_18_23} shows the fidelity averaged over the reference states as a function of $\kappa_{\rm ex}$.
In order to decrease numerical simulation time, we assume that the diagonal elements of the density matrix at $t=t_f$ are well approximated by the ones at $t=T_g$. This assumption is valid when the fidelity of the gate operations is high because the change of the diagonal elements during the ramping of the drive field and the free evolution after the ramping is negligible as exemplified in Fig.~\ref{fid_exp}(b).
The results obtained with this approximation are almost the same as the one without the approximation.
It is seen that the fidelity increases with the decrease of $\kappa_{\rm ex}$ because of the mitigation of unwanted effects of the decoherence.

\begin{figure}[h]
  \centering
  \includegraphics[width=7cm]{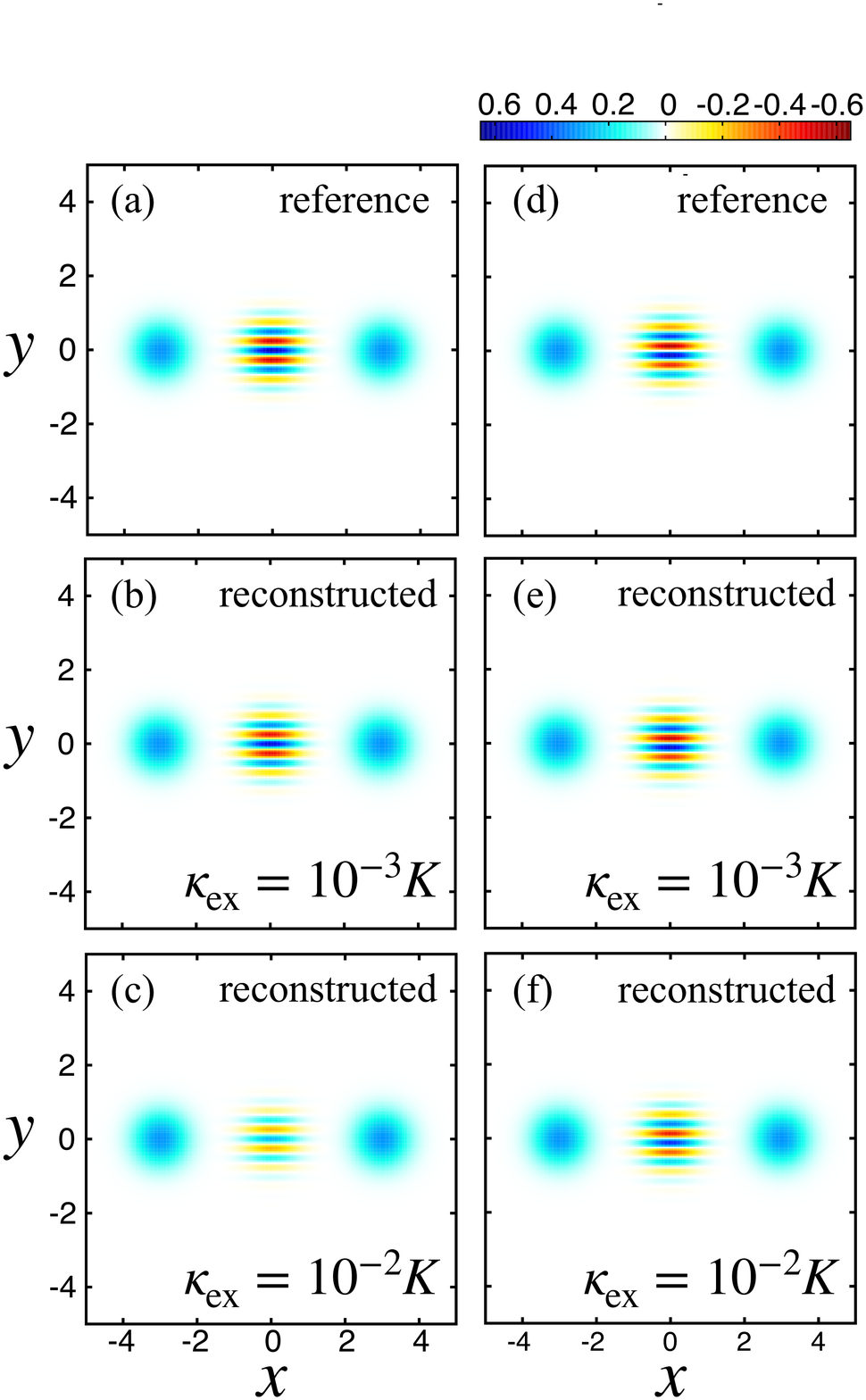}
  \caption{
Wigner function of the reference states $\rho_{x+,y+}$~(a,d) and their reconstructed states $\rho_{x+,y+}'$ for $\kappa_{\rm ex}=10^{-3}$~(b,e) and $10^{-2}$~(c,f).
The other used parameters are the same as Fig.~\ref{Fidelity_Rxy_3_18_23}.
}
 \label{Wig_3_19_23}
\end{figure}

Figure~\ref{Wig_3_19_23} exhibits the Wigner function of reference states $\rho_{x+}$, $\rho_{y+}$ and their reconstructed states.
The approximation used for the solid curve in Fig.~\ref{Fidelity_Rxy_3_18_23} is not used for this result and hereafter.
The Wigner function of the reconstructed states are approximately the same as those of the reference states for $\kappa_{\rm ex}/K=10^{-3}$.
On the other hand, the central fringe, which is the interference between the two stable coherent states and manifests the coherence of the KPO, becomes vague due to the decoherence during the gate operations for $\kappa_{\rm ex}/K=10^{-2}$.

\begin{figure}[h]
  \centering
  \includegraphics[width=7.5cm]{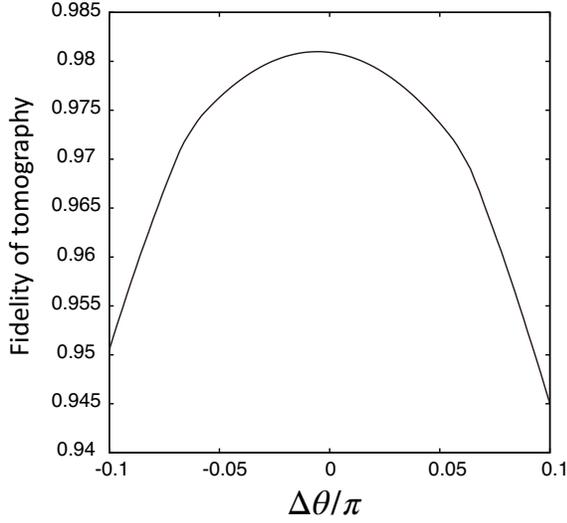}
  \caption{
Fidelity of the tomography as a function of $\Delta \theta$ for $\kappa_{\rm ex}/K=10^{-3}$.
The fidelity is averaged over the six different reference states.
The other used parameters are the same as Fig.~\ref{Fidelity_Rxy_3_18_23}.
}
 \label{rho1_3_19_23}
\end{figure}

We examine the sensitivity of the tomography to errors
in diagonal elements of the density matrix.
We assume that the measured value of the diagonal element corresponding to $|\tilde{0}\rangle\langle\tilde{0}|$ is given by $\cos^2[(\theta_{\rm true} + \Delta\theta)/2]$, while the true value is given by $\cos^2(\theta_{\rm true}/2)$.
Here, $\Delta\theta$ characterize the degree of the error.
The error influences not only the diagonal elements of the reconstructed density matrix $\rho'$ but also its off-diagonal elements because the reflection measurement is used for both of them.
We define  $\rho'_{11}$ by $\rho'_{11}=1-\rho'_{00}$ to satisfy ${\rm Tr}[\rho']=1$.
Because a density matrix should be positive semidefinite, we multiply the off-diagonal elements of $\rho'$ by $\eta(<1)$ if $\rho'$ is not positive semidefinite,
where $\eta$ is the maximum value that makes $\rho'$ positive semidefinite.
Figure~\ref{rho1_3_19_23} shows the fidelity averaged over the six different reference states as a function of $\Delta \theta$. Monotonic decrease is observed as $|\Delta\theta|$ increases in the parameter regime studied. The average fidelity is higher than 0.94 for $-0.1<\Delta \theta/\pi<0.1$ for the parameters used.

\section{Reflection coefficient without a drive field}
\label{Reflection coefficient without a drive field}
In this paper, a drive field is used to make the reflection coefficient dependent on the diagonal elements of the density matrix $\rho_{mm}(\tau)$.
In this section, we prove that the reflection coefficient is independent of $\rho_{mm}(\tau)$ without the drive field.

Without the drive field, the Hamiltonian of the KPO is written as Eq.~(\ref{Hamiiltonian_not-Omega}), 
and its eigenstates $\ket{\psi_m}$ have even or odd parity because the Hamiltonian is parity preserving.
The density operator $\rho(t)$ for $t\ge \tau+t_{\rm ramp} + t_{\rm delay}$ can be approximated by
\begin{equation}
\begin{split}
\rho'
&=
\rho_{00}(\tau)\ket{\tilde{0}}\bra{\tilde{0}}+\rho_{11}(\tau)\ket{\tilde{1}}\bra{\tilde{1}}
\\
&=
\frac{1}{2}\ket{\psi_0}\bra{\psi_0}
+
\frac{\rho_{00}-\rho_{11}}{2}\ket{\psi_0}\bra{\psi_1}\\
&+
\frac{\rho_{00}-\rho_{11}}{2}\ket{\psi_1}\bra{\psi_0}
+
\frac{1}{2}\ket{\psi_1}\bra{\psi_1},
\end{split}
\label{after_relaxation_10_23_22}
\end{equation}
with $\ket{\psi_{0,1}}=(\ket{\tilde{0}}\pm\ket{\tilde{1}})/\sqrt{2}$.
It is seen from Eq.~(\ref{after_relaxation_10_23_22}) that the information of $\rho_{00}$ is embedded in $\rho_{10}^{\rm (F)}[0]$ and $\rho_{01}^{\rm (F)}[0]$. Therefore, the reflection coefficient $\Gamma$ should depend on $\rho_{10}^{\rm (F)}[0]$ or $\rho_{01}^{\rm (F)}[0]$ for extraction of $\rho_{00}$.
However, $\Gamma$ is independent of  $\rho_{10}^{\rm (F)}[0]$ and $\rho_{01}^{\rm (F)}[0]$ because as seen in Eqs.~(\ref{reflection_coefficient}) and (\ref{Gamma_mn}) they appear in the reflection coefficient as a product with $X_{ij}=\braket{\psi_i|\hat{a}|\psi_j}$=0, where $i$ and $j$ have the same parity.
Note that $X_{ij}$ is zero when $\ket{\psi_i}$ and $\ket{\psi_j}$ have the same parity.

\section{Effect of off-diagonal elements of the density matrix}
\label{Effect of off-diagonal elements of the density matrix}
As explained in the main text, there is a one-to-one correspondence between $\rho_{00}$ and the reflection coeffiecient $\Gamma$ when off-diagonal elements of the density matrix are vanishing.
In order to realize the one-to-one correspondence, we set the delay time long enough so that the off-diagonal elements vanish in the main text.
However, as shown in this section, the effect of the off-diagonal elements to the reflection coefficient is negligible when the pump amplitude is sufficiently large. 
Therefore, the delay time can be set to zero in such a parameter regime.

\begin{figure}[]
  \centering
  \includegraphics[width=8.6cm]{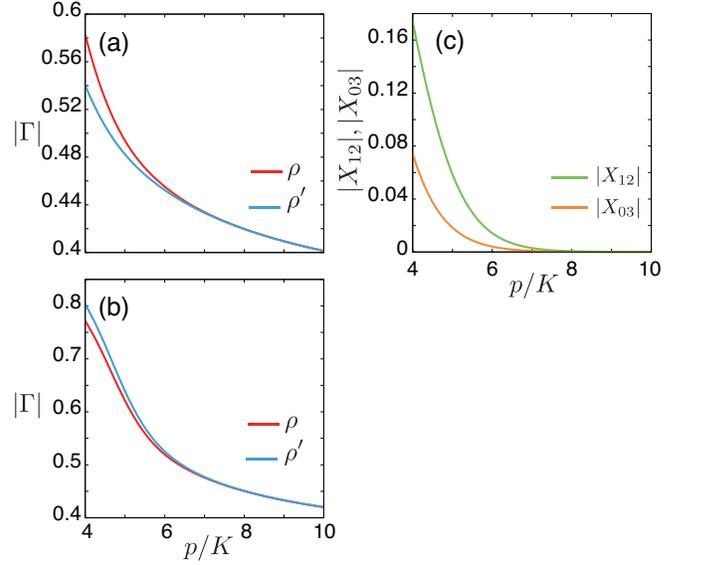}
  \caption{
The $p$ dependence of the amplitude of $\Gamma$ corresponding to $\rho$ (red) and $\rho'$ (blue) for $\omega_{\rm in}-\omega_p/2=\Delta \omega_{20}$ (a) and $\Delta \omega_{31}$ (b).
The $p$ dependence of $|X_{12}|$ (green curve) and $|X_{03}|$ (orange curve).
The used parameters are $\Omega/K=0.5$, $\kappa_{\rm ex}/K=0.01$ and $\kappa_{\rm in}/\kappa_{\rm ex}=0.5$.
}
 \label{off-diag}
\end{figure}

In order to examine the effect of the off-diagonal elements, we numerically calculate $\Gamma$ for the following two states with and without off-diagonal elements:
\begin{eqnarray}
\rho &=&[(\ket{\tilde{0}}+\ket{\tilde{1}})(\bra{\tilde{0}}+\bra{\tilde{1}})]/2,\nonumber\\
\rho' &=&[\ket{\tilde{0}}\bra{\tilde{0}} +\ket{\tilde{1}}\bra{\tilde{1}}]/2.
\label{}
\end{eqnarray}
Figures~\ref{off-diag}(a) and~\ref{off-diag}(b) show the amplitude of the reflection coefficient for $\omega_{\rm in}-\omega_p/2=\Delta \omega_{20}$ and $\Delta \omega_{31}$, respectively.
The difference between the reflection coefficients for $\rho$ and $\rho'$ becomes small when the pump amplitude $p$ increases.
This implies that the effect of the off-diagonal elements becomes negligible for the large-$p$ regime.

The insensitivity of $\Gamma$ to the off-diagonal elements comes from the fact that 
$X_{12}$ and $X_{03}$ become small in the large-$p$ regime as shown in Fig.~\ref{off-diag}(c).
The off-diagonal elements $\rho_{10}^{\rm (F)}[0]$ and $\rho_{01}^{\rm (F)}[0]$ appear
in the form of $X_{12}^\ast \rho_{10}^{\rm (F)}[0]$ and $X_{03}^\ast \rho_{01}^{\rm (F)}[0]$ in the reflection coefficient in Eq.~(\ref{reflection_coefficient}).
Therefore, the effect of the off-diagonal elements becomes negligible when $X_{12}$ and $X_{03}$ are sufficiently small.

The decrease of $|X_{12}|$ and $|X_{03}|$ in the large-$p$ regime is explained as follows.
As $p$ increases, the four highest eigenstates of Hamiltonian (\ref{Hamiiltonian_Omega}) can be approximated as $\ket{\psi_{0(2)}}=D(\alpha)\ket{0(1)}$ and $\ket{\psi_{1(3)}}=D(-\alpha)\ket{0(1)}$.
Then, we have
\begin{equation}
\begin{split}
X_{12}=X_{03}=e^{-2|\alpha|^2}(1-2|\alpha|^2),
\end{split}
\label{X12_11_25_22}
\end{equation}
where $\alpha=\sqrt{p/K}$.
From Eq.~(\ref{X12_11_25_22}), it is seen that $|X_{12}|$ and $|X_{03}|$ become small when $p$ is sufficiently large. 

\section{Energy eigenstates and their asymptotic form}
\label{Energy eigenstates}
As mentioned in the main text, the eigenstates of $H_{\rm KPO}$ in Eq.~(\ref{Hamiiltonian_Omega}) can be approximated by $D(\pm\alpha)\ket{m}$.
Here, we quantitatively examine the validity of the approximation.

\begin{figure}[]
  \centering
  \includegraphics[width=8.6cm]{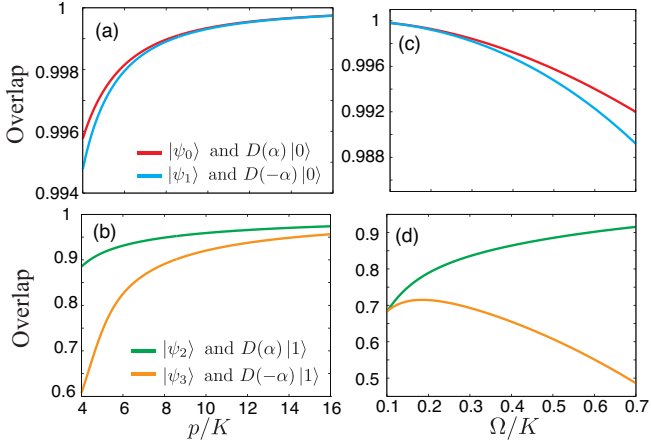}
  \caption{
Overlap between energy eigenstates and their approximations $D(\pm\alpha)\ket{m}$ is shown as a function of $p$ (a,b) and $\Omega$ (c,d).
The used parameters are $\kappa_{\rm ex}/K=0.01$, $\kappa_{\rm in}/\kappa_{\rm ex}=0.5$,  $\Omega/K=0.5$~(a,b) and $p/K=4.0$~(c,d).
}
 \label{fid_gamma}
\end{figure}

Figure~\ref{fid_gamma} shows the overlap, $|\bra{\psi_i}{\psi_i'}\rangle|^2$, between relevant energy eigenstates $\ket{\psi_i}$ and their approximated one $\ket{\psi_i'}=D(\pm\alpha)\ket{m}$ as a function of $p$ [Figs.~\ref{fid_gamma}(a) and \ref{fid_gamma}(b)] and also as a function of $\Omega$ [Figs.~\ref{fid_gamma}(c) and \ref{fid_gamma}(d)].
It is seen that the approximation becomes more valid when $p$ increases as seen in Figs.~\ref{fid_gamma}(a) and \ref{fid_gamma}(b).

The validity of the approximation can be degraded as $\Omega$ increases.
The overlap for $\ket{\psi_{0,1}}$ is high for the small $\Omega$ regime as seen in Fig.~\ref{fid_gamma}(c).
However, the overlap is decreased as $\Omega$ becomes large.
We consider that this is due to the distortion of the potential under the strong drive field.
The overlap for $\ket{\psi_{2,3}}$ is low in the small $\Omega$ regime as seen in Fig.~\ref{fid_gamma}(d).
This is because that these states are loosely trapped by the double-well potential and their Winger function is distributed among both the wells when $\Omega$ is small. 
On the other hand, as $\Omega$ is increased, $\ket{\psi_{2,3}}$ is trapped in either of the wells, and therefore the overlap increases.
However, because $\ket{\psi_3}$ is confined loosely in the well, the overlap starts to drop due to the distortion of the potential when the drive field is further strengthened. 

\section{Pure dephasing}
\label{Pure dephasiing}
We consider the case that there is pure dephasing with the rate of $\gamma$.
The pure dephasing enhances the relaxation of the diagonal elements of the density matrix, and increases the population of excited states out of the qubit subspace.

To examine the effect of the pure dephasing on the density matrix, we numerically solve the master equation
\begin{equation}
\begin{split}
\frac{d\rho(t)}{dt}
&=
-\frac{i}{\hbar}
\left[H(t),\rho(t)\right]
+
\frac{\kappa_{\rm tot}}{2} \mathcal{D}[\hat{a}]\rho(t)
+
\gamma \mathcal{D}[\hat{a}^\dagger \hat{a}]\rho(t),
\end{split}
\label{master eq_gamma}
\end{equation}
with $H(t)=-\frac{K}{2}\hat{a}^{\dagger 2} \hat{a}^2+\frac{p}{2}\left(\hat{a}^2+\hat{a}^{\dagger 2}\right)+ \Omega(t)(\hat{a}^{\dagger} + \hat{a})$,
where $\Omega(t)$ is given by Eq.~(\ref{Omega_11_15_22}).
In the numerical simulation, the initial state is set to be $\rho(0)=(\sqrt{0.2}\ket{\tilde{0}} + \sqrt{0.8}\ket{\tilde{1}})(\sqrt{0.2}\bra{\tilde{0}} + \sqrt{0.8}\bra{\tilde{1}})$. 

\begin{figure}[]
  \centering
  \includegraphics[width=8.6cm]{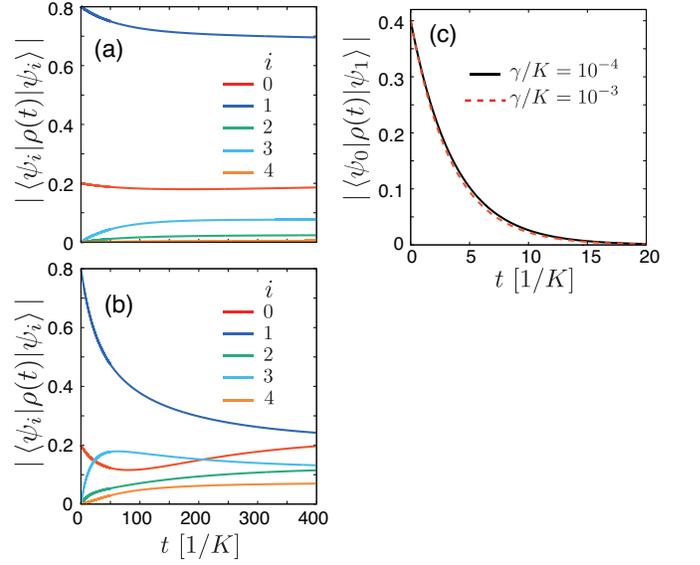}
  \caption{
Time dependence of the population of relevant five levels, $\braket{\psi_i|\rho(t)|\psi_i}$ for $\gamma/K=10^{-4}$~(a) and $10^{-3}$~(b).
 The red, blue, green, light blue and orange curves are for $i=0,1,2,3$ and 4, respectively.
The black solid and red dashed curves in panel (c) represent the off-diagonal element $\braket{\psi_0|\rho(t)|\psi_1}$ for $\gamma/K=10^{-4}$ and $10^{-3}$, respectively.
The used parameters are $p/K=9.0$, $\Omega_0/K=0.1$, $t_{\rm ramp}=20/K$, 
$\kappa_{\rm ex}/K=10^{-2}$ and $\kappa_{\rm in}/\kappa_{\rm ex}=0.5$.
  }
 \label{kappa_ex_10_21_22}
\end{figure}

The population of the first five levels defined by $\braket{\psi_i|\rho(t)|\psi_i}$ are exhibited in Fig.~\ref{kappa_ex_10_21_22}.
The change of the population is much faster than the case without pure dephasing shown in Fig.~\ref{fid_exp}(b).
The energy levels out of the qubit subspace are also populated due to the relaxation caused by the pure dephasing.
It is also seen that the larger $\gamma$ is, the larger the change of the populations are. 

\begin{figure}[]
  \centering
  \includegraphics[width=7cm]{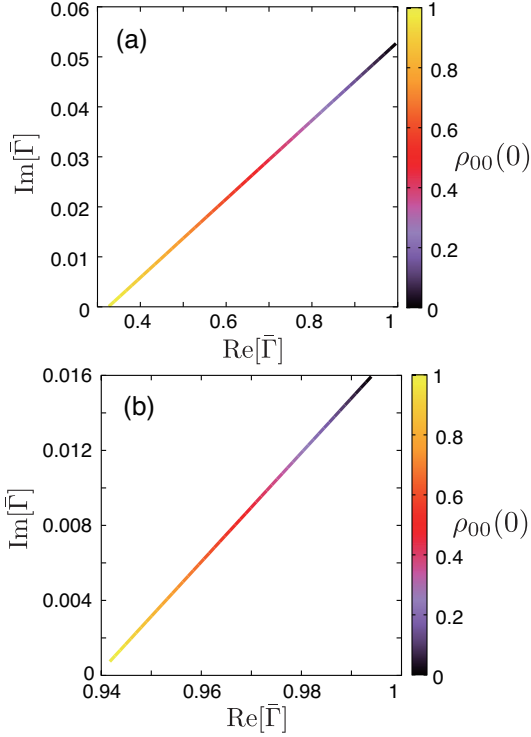}
  \caption{
The relationship between the time-averaged reflection coefficient $\bar{\Gamma}$ and the diagonal element at the initial time $\rho_{00}(0)$ for $\gamma/K=10^{-4}$ (a) and $10^{-3}$ (b).
The used parameters are $p/K=9.0$, $\Omega_0/K=0.1$, $t_{\rm ramp}=20/K$, $\kappa_{\rm ex}/K=10^{-2}$ and $\kappa_{\rm in}/\kappa_{\rm ex}=0.5$.
}
 \label{time-ave_Gamma_11_9_22}
\end{figure}

This result implies that  the reflection measurement should be performed before the KPO relaxes to its stationary state.
The reflection coefficient can be derived for the case with pure dephasing in the same manner as Ref.~\cite{Masuda2021b}.
The reflection coefficient can be written as Eq.~\ref{reflection_coefficient} with
\begin{equation}
\begin{split}
\xi_{mn}
=
\frac
{\kappa_{\rm ex} X_{mn}\sum_{k}\left(X^*_{kn}\rho_{km}^{\rm (F)}[0]-\rho_{nk}^{\rm (F)}[0] X^*_{mk}\right)}
{
i\Delta_{nm}
+\kappa_{\rm tot}X_{nn}X^*_{mm}
-\frac{\kappa_{\rm tot}}{2}\left(Y_{nn}+Y_{mm}\right)
+L
},
\\
\end{split}
\label{reflection_coefficient_gamma}
\end{equation}
where $L=2\gamma Y_{nn} Y^*_{mm}
-\gamma \left(Z_{nn}+Z_{mm}\right)$ and $Z_{mm}=\bra{\psi_m} (a^\dagger a)^2 \ket{\psi_m}$.
We assume that the reflection measurement is performed for $20/K \le t \le 220/K$ and 
the time averaged reflection coefficient $\bar{\Gamma}$ is obtained. 
This duration of the measurement is approximately 3 $\mu$s for $K/2\pi=10$ MHz, which is experimentally feasible~\cite{Yamaji2021}. 
The initial state is set to be $\rho(0)=(\sqrt{\rho_{00}(0)}\ket{\tilde{0}}+\sqrt{1-\rho_{00}(0)}\ket{\tilde{1}})(\sqrt{\rho_{00}(0)}\bra{\tilde{0}}+\sqrt{1-\rho_{00}(0)}\bra{\tilde{1}})$.
Figure~\ref{time-ave_Gamma_11_9_22} shows $\bar{\Gamma}$ as a function of $\rho_{00}(0)$.
It is seen that $\bar{\Gamma}$ monotonically changes with $\rho_{00}(0)$.
Therefore, we can extract  $\rho_{00}(0)$ from $\bar{\Gamma}$.

When the input field is resonant to a transition $\ket{\psi_m}\rightarrow\ket{\psi_n}$ and off-resonant to other transitions, the reflection coefficient can be written as $\Gamma=1+\xi_{mn}$.
On the other hand, the reflection coefficient of a linear resonator is written as
\begin{equation}
\begin{split}
\Gamma_{\rm r}=1+\frac{\kappa^{\rm (r)}_{\rm ex}}{i\Delta_{\rm r}-(\kappa^{\rm(r)}_{\rm ex}+\kappa^{\rm (r)}_{\rm int})/2},
\end{split}
\label{Gamma_res_10_23_22}
\end{equation}
with the external decay rate $\kappa^{\rm (r)}_{\rm ex}$, internal decay rate $\kappa^{\rm (r)}_{\rm int}$ and detuning $\Delta_{\rm r}=\omega_{\rm in}-\omega_{0}$ where $\omega_0$ is the angular resonance frequency of the resonator.
Comparing $\Gamma=1+\xi_{mn}$ with Eq.~(\ref{Gamma_res_10_23_22}), the nominal external and internal decay rates for the KPO~\cite{Masuda2021b} can be defined as
\begin{align}
\begin{split}
&\tilde{\kappa}^{(mn)}_{\rm ex}=\kappa_{\rm ex} X_{mn}\sum_{k}\left(X^*_{kn}\rho_{km}^{\rm (F)}[0]-\rho_{nk}^{\rm (F)}[0] X^*_{mk}\right),
\\
&\tilde{\kappa}^{(mn)}_{\rm int}=
-2\kappa_{\rm tot}X_{nn}X^*_{mm}
+\kappa_{\rm tot}\left(Y_{nn}+Y_{mm}\right)\\
&\hspace{35.5pt}-4\gamma Y_{nn} Y^*_{mm}
+2\gamma \left(Z_{nn}+Z_{mm}\right)\\
&\hspace{35.5pt}-\kappa_{\rm ex} X_{mn}\sum_{k}\left(X^*_{kn}\rho_{km}^{\rm (F)}[0]-\rho_{nk}^{\rm (F)}[0] X^*_{mk}\right).
\end{split}
\end{align}

\begin{figure}[h!]
  \centering
  \includegraphics[width=6.5cm]{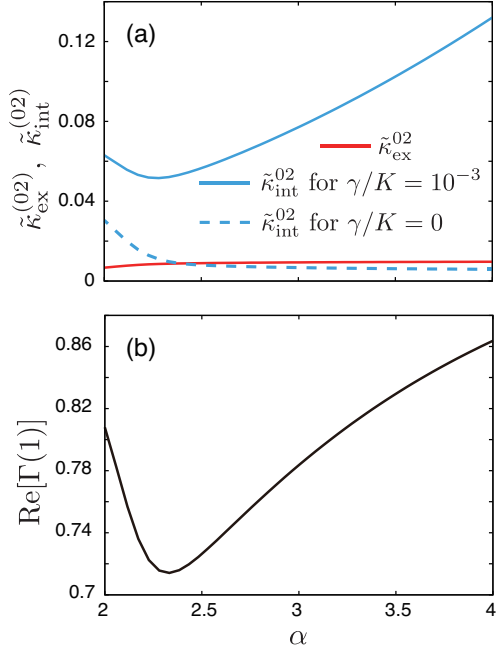}
  \caption{
Nominal internal and external decay rates (a) and the real part of the reflection coefficient for $\rho_{00}=1$ (b) as functions of $\alpha (=\sqrt{p/K})$.
The blue and red curves represent the nominal internal and external decay rates for $\gamma/K=10^{-3}$, respectively, in panel (a).
In panel (a), the blue dashed curve is the nominal internal for $\gamma/K=0$.
The other parameters are $p/K=9.0$, $\Omega/K=0.1$,
$\omega_{\rm in}-\omega_p/2=\Delta\omega_{20}$, 
 $\kappa_{\rm ex}/K=10^{-2}$ and $\kappa_{\rm in}/\kappa_{\rm ex}=0.5$.
 \label{nominal_decay_11_08_22}
}
\end{figure}

Figure~\ref{nominal_decay_11_08_22}(a) shows the nominal internal and external decay rates as a function of $\alpha$.
The frequency of the input field is set to the one corresponding to the transition from $\ket{\psi_0}$ to $\ket{\psi_2}$.
The nominal internal decay rate is much higher than the nominal external decay rate for $\gamma/K=10^{-3}$, that is, the KPO is in the under-coupling regime.
Therefore, the reflection coefficient is insensitive to the frequency of the input field compared to the case without the pure dephasing.

As seen in Fig.~\ref{nominal_decay_11_08_22}(a), the difference between the nominal internal and external decay rates becomes small at $\alpha\sim2.3$.
At this point, the sensitivity of the reflection coefficient to $\rho_{00}$ becomes large compared to other points as indicated by $\rm{Re}[\Gamma(1)]$ in Fig.~\ref{nominal_decay_11_08_22}(b).
(Note that $\Gamma(0)\simeq 1$ in the range of $\alpha$ used for this figure.)
This result implies that there is a suitable point of $\alpha$ for the measurement of a KPO along the $z$-axis.


\end{document}